\def \be {\begin{equation}}
\def \eq {\end{equation}}
\def \bee {\begin{eqnarray}}
\def \eqq {\end{eqnarray}}

\def \bea {\begin{array}{c}}
\def \eqa {\end{array}}

\def \la {\langle}
\def \ra {\rangle}
\def \R {{\bf R}}
\def \C {{\bf C}}
\def \Z {{\bf Z}}
\def \del {\partial}
\def \dels {\partial\kern-.5em / \kern.5em}
\def \As {{A\kern-.5em / \kern.5em}}
\def \Ds {D\kern-.7em / \kern.5em}

\def \a {\alpha}
\def \b {\beta}

\def \G {\Gamma}
\def \d {\delta}
\def \eps {\epsilon}

\def \s {\sigma}

\def \one {{\bf 1}}

\def \H {{\cal H}}
\def \A {{\cal A}}
\def \O {{\cal O}}
\def \Gc {{\cal G}}

\def \Rt {\tilde{R}}
\def \dt {\tilde{d}}
\def \Ut {\tilde{U}}
\def \Ub {{\bf U}}
\def \Ubt {\tilde{\Ub}}

\documentstyle[12pt]{article}

\begin{document}

\begin{titlepage}
\today          \hfill
\begin{center}
\hfill    UU-HEP/98-01\\
\hfill    \\

\vskip .5in

{\large Noncommutative Gauge Theories
in Matrix Theory}

\vskip .5in

Pei-Ming Ho$^{\dagger}$
\footnote{Present address: Department of Physics,
Jadwin Hall, Princeton University, Princeton, NJ 08544.}
and Yong-Shi Wu$^{\ddagger}$
\footnote{On sabbatical from Department of Physics,
University of Utah, Salt Lake City, UT 84112-0830.}
\vskip .3in
{\em $^\dagger$ Department of Physics,
University of Utah \\
Salt Lake City, Utah 84112-0830}
\vskip .3in
{\em $^\ddagger$ School of Natural Sciences,
Institue for Advanced Study\\
Princeton, New Jersey 08540}

\end{center}

\vskip .5in

\begin{abstract}
We present a general framework for Matrix
theory compactified on a quotient
space ${\bf R}^n/\G$, with $\G$
a discrete group of Euclidean motions in
${\bf R}^n$. The general solution to
the quotient conditions gives a gauge
theory on a noncommutative space. We
characterize the resulting noncommutative
gauge theory in terms of the twisted group
algebra of $\G$ associated with a projective
regular representation. Also we show how to
extend our treatments to incorporate
orientifolds. 
\end{abstract}

\end{titlepage}
\newpage
\renewcommand{\thepage}{\arabic{page}}
\setcounter{page}{1}

\section{Introduction}

According to general relativity,
classical gravity is nothing but (spacetime)
geometry. It has long been suspected that
quantum fluctuations of gravity near the Planck
scale may make points in space fuzzy and,
therefore, call for modifying our current
description of geometry. Recent progress
in string theory has indeed confirmed
this suspicion.
Facts and arguments on the lately
discovered string dualities have pointed
to the existence of a fundamental
quantum theory in eleven dimensional
spacetime, called M theory, which
underlies all known five perturbative
superstring theories \cite{MT}. The
BFSS matrix model was proposed in
ref. \cite{BFSS} for the microscopic
description of M theory in discrete
light-cone quantization \cite{Suss},
in terms of a set of $N$ partons,
called D0-branes, on which strings can end.
A novel feature of the M(atrix) theory
is that the nine transverse coordinates
$X^\mu\; (\mu=1,2,...9)$ of the D0-branes
are promoted \cite{Witten} into
$N\times N$ Hermitian matrices.
One smells the need for new geometry
that deals with spaces whose coordinates
(as functions on the space)
are noncommutative. Such spaces are
called {\it noncommutative spaces},
and their geometry {\it noncommutative
geometry}
(NCG).

NCG, as Connes advocated \cite{Conn}, deals with
a geometric space {\it not} as a set of points,
instead starting with the set of all functions
defined on it. For usual manifolds, the latter
forms a commutative algebra, from which one can
reconstruct the underlying manifold, in accordance
with the Gel'fand-Na\v{i}mark theorem \cite{Gelf}.
But in NCG, it can be a noncommutative algebra.
The precise data for defining a noncommutative
space consist of the spectral triple
$({\cal A},{\cal H},{\cal D})$ \cite{Conn}.
Here ${\cal A}$ is an associative
algebra, thought of as the algebra of functions
(including the coordinates) on the space.
${\cal H}$ is a Hilbert space that represents
the algebra ${\cal A}$ as operators acting on
it, thought of as the Hilbert space on which
the coordinates are represented as operators;
the trace associated with the inner product
of ${\cal H}$ can be used to define the
integration on the space. Finally ${\cal D}$
is a derivation acting on ${\cal H}$, called
the Dirac operator, representing the usual
derivative operator.

Gauge theory on a noncommutative space will be
abbreviated as {\it noncommutative gauge
theory} (NCGT). The gauge group $\Gc(\A)$
is the group of unitary elements in $\A$,
while the covariant derivative is obtained
by adding terms of the form
$\sum_i a_i[D, b_i]$ for $a_i, b_i\in\A$
to the Dirac operator. The generalized gauge
potential in this form is used to incorporate the
usual Higgs fields \cite{CoLo}
in the Yang-Mills-Connes
action. Previously we have shown \cite{HW}
that the BFSS matrix model action, which
is given by the dimensional reduction of ten
dimensional supersymmetric $U(N)$ Yang-Mills
theory down to $0+1$ dimension, can be
understood as an NCGT: The defining
algebra is ${\cal A}_0= M_N(\C)$, that
of $N\times N$ complex matrices, and
the Hilbert space is ${\cal H}_0=\C^N$.
The Dirac operator is simply
${\cal D}=\Gamma^0 (\del_0 + A_0) +
\Gamma^\mu X_\mu$. Here both the
gauge potential $A_0$ and the
``Higgs'' fields $X^\mu$
$(\mu=1,2....,9)$ are the generalized
gauge potentials.
(In the following the explicit form of
the BFSS action is not needed.)

In a recent paper\cite{CDS}, Matrix
theory on a torus is shown to be
described by NCGT on a quantum
torus. A further case-by-case study
is given in ref. \cite{HWW}. In
this note we will show that Matrix
theory compactified on ${\bf R}^n/\G$,
with $\G$ a discrete group of
Euclidean motions in ${\bf R}^n$,
generically leads to NCGT
characterized by the group algebra of
$\G$ twisted by a projective regular
representation. Appropriate
$\bf{Z}_2$-grading or $\bf{Z}_2$-graded
extension of $\Gamma$ will incorporate
orientifolds.

\section{Quotient Conditions}

We want to study the compactification
of some transverse directions on a flat
quotient space ${\bf R}^n/\G$, with
$n\leq 9$ and $\G$ a discrete group
of the Euclidean motions in ${\bf R}^n$.
If the action of $\G$ is free (i.e. has
no fixed points), then ${\bf R}^n/\G$ is a
manifold with $\G$ as the fundamental
group; otherwise it is an orbifold.

For $g\in \G$, we write the action
of $g$ on ${\bf x}\in {\bf R}^n$
as ${\bf x} \to {\bf x}g \equiv
R_g ({\bf x}) + T_g ({\bf x})$,
where $R_g$ is a rotation, while $T_g$
a translation. For simplicity, assume
$\G$ preserves orientation and
consider the naturally lifted action
of $\G$ on the matrix-valued $X^{\mu}$,
denoted as $\Phi_g$ for $g\in \G$:
\be
\Phi_g^{\mu}(X)
= (R_g)^{\mu}_{\nu} X^{\nu}
 + d_g^{\mu}\; {\bf I}.
\eq
$X^\mu$ is unchanged if it is not in the
compactified directions. The superpartner
$\Psi$ transforms under $\Phi_g$ as a
ten-dimensional Majorana-Weyl spinor
under the proper rotation $R_g$.
Or one may work out the action of
$\Phi_g$ on $\Psi$ by requiring the
Matrix model action be invariant.
Below we will concentrate on the bosonic
variables, since the fermionic ones
can be similarly treated.

To implement the compactification,
we follow the techniques for dealing
with D-branes on a quotient space
\cite{GP,DM,Tay}. Namely for a
D0-brane located at some point
in ${\bf R}^n/\G$, we need to
consider all image D0-branes in
${\bf R}^n$ under the action of
$\G$, locating on a $\G$-orbit.
Then the (Chan-Paton) label
for the D0-brane is extended
from a single $i (=1,2,...,N)$ to
a pair $(g,i)$ with $g\in\G$.
The compactification (or quotient)

to $\R^n/\G$ implies
gauging the discrete symmetry $\G$
for the D0-brane quantum mechanics,
or the gauge equivalence of the
open strings described by the
coordinate matrix element
$(X^\mu)_{(g_1,i),(g_2,j)}$
and by its image under simultaneous
action of $\G$ on $g_1$ and $g_2$:
\be
(X^\mu)_{(g_1g,i),(g_2g,j)}
= \Phi_{g}^{\mu}
(X_{(g_1,i),(g_2,j)}).
\eq

We introduce a set of unitary operators
$\{U_g: g\in \G\}$ to implement
the action of $\G$ on
the matrix variables:
\be
U_g^{-1} X^{\mu} U_g
= \Phi_g^{\mu}(X).
\label{qc}
\eq
Then gauging the discrete symmetry
$\G$ can be achieved by 1) including
$U_g$'s into the theory and making
them part of the gauge group, so
that the physical states are invariant
under $\G$ and 2) extending path
integral quantization to include the
twisted sectors, which are represented
by the solutions to the above quotient
conditions (\ref{qc}). Note that the
shift operator $U_g$ also admits
the following interpretation
in string picture: Viewed from $\R^n$,
corresponding to each $U_g$
there is an open string stretching
between a D0-brane and one of its
images that is labelled by $g$. Upon
compactification to $\R^n/\G$, it
becomes a string (in the ground
state) winding on the 1-cycle
corresponding to $g$. As the size of
$\R^n/\G$ tends to zero, these
winding states become massless, so we
have to incorporate them into the
compactified theory.

\section{Projective Representation and
Twisted Group Algebra}

It follows from the group property of
$\Phi_g$ in the conditions (\ref{qc})
that the action of $U_g U_h$ is the same
as that of $U_{gh}$, so they can differ
only by a phase factor:
\be
U_g\,U_h = q(g,h)\; U_{gh}\, ,
\label{proj}
\eq
with $q(g,h)=\exp \{ i \alpha (g,h)\}$.
Here $q(g,h)$ or $\a(g,h)$ depends on
a pair of group elements $(g,h)$.
We do not want to impose constraints
more than necessary \cite{comm1}, the
operator $U_{e}$  (corresponding to the
identity $e$ of $\G$) has to be the
identity operator $\one$ up to a phase
factor. Without loss of generality we
rescale $U_e$ to $\one$.
Then it follows from eq. (\ref{proj})
that $q (g,e)= q (e,g)=1$.
The associativity
$(U_f\,U_g)\,U_h= U_f\,(U_g\,U_h)$
leads to the 2-cocycle condition:
\bee
q(f,g) q(fg,h)= q(f,gh) q(g,h).
\label{2coc}
\eqq
Thus, the operators $U_g$'s in the
quotient conditions form a {\it faithful,
projective} representation of $\G$,
determined by a {\it 2-cocycle} $q (g,h)$.
The faithfulness implies that only $U_e$
is proportional to $\one$. Physically we
need this condition, in order for the
quotient conditions to faithfully
describe the desired compactification.

For instance, if $g$ and $h$ commute
with each other: $gh=hg$, then the difference
$\theta (g,h)=\alpha (g,h)-\alpha (h,g)$
is a cohomological invariant. So the
projectivity condition (\ref{proj}) can
be replaced by
\be
U_g\,U_h =
\exp \{i\theta (g,h)\} U_h\, U_g.
\label{theta}
\eq

Using each $U_g \; (g\in \G)$ as a basis
vector, we can generate a vector
space with complex coefficients, whose
dimension is the order $|\G|$ of
the group $\G$, i.e. the number
of elements in $\G$, which is either
finite or countable. Upon introducing
multiplication of two $U$'s by
eq. (\ref{proj}), this complex vector space
is turned into an algebra, denoted as
$\C^{\alpha}\G$, called the group
algebra of $\G$ twisted (or deformed)
by the 2-cocycle $\alpha$.

Now we come to the key point of our
approach: In the spirit of NCG
using an algebra to define a space,
we use the twisted group algebra
$\C^{\alpha}\G$ to define a
noncommutative space, and construct
a Hilbert space ${\cal H}_\Gamma$ to
represent the algebra. It is natural
to take it to be the linear space spanned
by $\{U_g\}$ in the projective {\it regular}
representation: The $U_g$'s act on
$\C^{\alpha}\G$  by multiplication.
There is a one-to-one correspondence
between the basis $U_g$ in $\C^{\alpha}\G$
and the basis states in
${\cal H}_\Gamma$. The state
corresponding to the identity
operator $U_e$ is called the
``vacuum'' state, denoted
as $\ra$. Then the state
corresponding to $U_h$ is denoted
as $U_h\ra$. Now $U_g$'s
are represented as operators
on ${\cal H}_\Gamma$ whose
action is the same as their action
on $\C^{\alpha}\G$.

Moreover, we need to define an
inner product in ${\cal H}_\Gamma$,
which should make the operators
$U_g$ unitary. It is easy to see
that the inner product should be
defined by the cyclic linear functional
\be
\la U_g\ra
= \delta(g,e),
\label{inpr}
\eq
where $\delta(g,e)$ is $0$ if
$g\neq e$, and is $1$ if $g=e$.
Then the trace over $\H_\Gamma$ is simply
$|\G|$ times this linear functional.

\section{General Solution to Quotient Conditions}

Before solving the quotient conditions,
upon extending the Chan-Paton indices from
$i$ to $(g,i)$, the algebra $\A$ of the spectral
triple defining the Matrix model is enlarged
to $\A\equiv\O(\H_\G)\times \A_0$, where
$\O(\H_\G)$ is the algebra of operators
on $\H_\G$, while the gauge group is the group,
$\Gc (\A)$, of all unitary elements in
the algebra $\A$. Our problem of Matrix theory
compactification is now reduced to finding
the general solution to the quotient
conditions on the noncommutative space,
namely to write down the general solution
for $X^\mu$'s, which are understood as
operators in $\A$ acting on the Hilbert
space ${\cal H}\equiv\H_\G\times \H_0$.

To find the general solution, one may
follow Zumino's prescription \cite{HWW}.
The quotient conditions imply that the
action of $X$ on the basis of
${\cal H}_{\Gamma}$, consisting of
$U_h$'s acting on the ``vacuum''
state $\ra$, is determined by its
action on the vacuum, which can be
an arbitrary state in $\H$:
\be
X^\mu\ra = A^\mu(U)\ra.
\label{act0}
\eq
Here $A^\mu(U)=\sum_{g\in \G}a^\mu(g) U_g$
(with $a^\mu(g)\in \A_0$)
is a general element of the algebra $\A$.
Then for the state $X^\mu U_h\ra$,
one may use the quotient
conditions (\ref{qc})
to move $X^\mu$ to the right, then
use eq. (\ref{act0}) to obtain
\be
U_h \Phi_h^\mu  (X)\ra
= U_h [(R_h)^\mu_\nu A^\nu(U)+d_h^\mu ]\ra,
\label{8}
\eq
Introducing the projective operators
$P_g$ for $g\in \G$:
\be \label{Pg}
P_g U_h\ra=\d(g,h) U_h\ra\, ,
\eq
and the elements of some ``dual'' algebra
\be
\Ut_h\equiv \sum_{g\in\G}
U_g U_h U_g^{-1}P_g,
\label{Ut}
\eq
then eq. (\ref{8}) can be written as
\be
[ A^\nu(\Ut)\Rt^{\mu}_{\nu}
+\dt^\mu ] U_h\ra,
\eq
where
\be\label{Rtdt}
\Rt^\mu_\nu\equiv\sum_{g\in\G}(R_g)^\mu_\nu P_g,
\quad \dt^\mu\equiv\sum_{g\in\G}d_g^\mu P_g.
\eq
Thus the general solution of $X$ is
\be
X^\mu=A^\nu(\Ut)\Rt^\mu_\nu+\dt^\mu.
\label{gs}
\eq
All physical (gauge field) degrees of freedom in $X$
reside in the function $A^\mu(\Ut)$ defined on
the dual space, which can be viewed as the generalized
gauge field in NCGT.

As for $A_0$ and $X^\mu$'s not in the compactified
directions, they are invariant under $U_g$, so
the solutions are simply $A_0=A_0(\Ut)$
and $X^\mu=X^\mu(\Ut)$. (See eq. (\ref{18}) below.)

The constant operators $\Rt^\mu_\nu$ and $\dt^\mu$
commute among themselves and satisfy
\bee
&\Rt^\mu_\nu\Ut_g=\Ut_g\Rt^\mu_\s (R_g)^\s_{\nu},
\\
&\dt^\mu\Ut_g=\Ut_g(\Rt^\mu_\nu d^\nu_g+\dt^\mu).
\label{aU}
\eqq
{}For a group of translations,
$(R_g)^\mu_{\nu}=\d^\mu_\nu$ for all
$g\in\G$, so $\Rt^\mu_\nu=\d^\mu_\nu$
and eq. (\ref{aU}) suggests that $\dt^\mu$
are derivatives with respect to the
exponents of $\Ut$ \cite{CDS,HWW}. In
general, the operator $\Rt$ also has
the interpretation of a derivative on
a noncommutative space. A simple example
was utilized in \cite{CoLo} to formulate
the Higgs field in the standard model
as the covariant derivative on the
space of two points.

\section{The Resulting Noncommutative Gauge Theory}

To characterize the resulting
theory as NCGT, let us first note that after
imposing the quotient conditions, the surviving
group $\Gc'$ of (local) gauge symmetry
becomes the commutant of $\A_\G\equiv\C^{\alpha}\G$
in $\Gc(\A)$, i.e.,
\be
\Gc'=\{g\in\Gc(\A): [g, U_h]=0, \,\forall h\in\G\}.
\eq
Hence one may take the algebra in the spectral
triple defining the compactified Matrix model
to be the commutant of $\A_\G$ in $\A$:
\be
\A'=\{a\in\A: [a, U_h]=0, \,\forall h\in\G\},
\eq
so that $\Gc'$ is the group of unitary elements in $\A'$.

{}From the general solution (\ref{gs}), it is
easy to see that $\A'= \A'_\G\times \A_0$, where
$\A'_\G$ is spanned by the operators $\Ut_g$'s.
It is easy to verify that
\bee
&&[\Ut_h, U_g]=0, \quad \forall h,g\in\G,
\label{18}\\
&& \Ut_g\Ut_h=e^{i\a(h,g)}\, \Ut_{hg}.
\label{19}
\eqq
Thus $\A'_\G$ is isomorphic to the algebra
obtained from $\A_\G$ by reversing the
ordering of all products. There is also
a one-to-one correspondence between $\H_\G$
and $\A'_\G$ given by $\Ut_g\ra = U_g \ra
\leftrightarrow \Ut_g$.

In the spirit of NCG, we use the algebra
$\A'_\G$ and the associated Hilbert space
${\cal H}_\Gamma$ to define the noncommutative
dual space for the compactified Matrix
model. Both of them are characterized
by a projective (including genuine) regular
representation of $\G$, known to be faithful.

Consider the group of elements $g\in\Gc(\A)$
preserving the quotient conditions, i.e.
\be \label{gUg}
gU_h g^{-1}=e^{i\b(h)}U_h\, ,
\eq
for some $\b(h)$ for all $h\in\G$.
The gauge group $\Gc'$ is the subgroup of
those elements with $\b(h)=0$ for all $h\in\G$.
It is easy to see that $\b(h)$ has to be a
1-cocycle for (\ref{gUg}) to be consistent
with $\A_\G$. There is a one-to-one
correspondence between $H^1(\G,U(1))$ and
scalings of $U_h$ by phase factors which
can be realized as conjugation by elements
in $\Gc(\A)$. Such transformations shift
the exponents of $\Ut_h$ (and $U_h$) by
constants, so $H^1(\G,U(1))$ can be viewed
as the global symmetry group of translations
on the dual space. Generally any algebra
automorphism of $\A'_\G$ is a global
symmetry not existing before compactification.

Substituting the solution (\ref{gs})
into the BFSS action, we will get the
(bosonic part of) action for the
resulting NCGT, including
deformed Yang-Mills theories and
gauged sigma models (see below
for examples).

\section{Examples}

\subsection{Matrix Theory on Quantum Tori}

To show how our abstract approach
works in practice, let us first
examine the case when $\G$ is
generated by $n$ translations
$t_a$ along $d_a^\mu$
($\mu,a=1,...,n\geq 2$).
Since $\G$ is abelian, eq.
(\ref{theta}) applies. Taking
$g=t_a, h=t_b$, all nontrivial
2-cocycles of $\G$ are
determined by $\theta_{ab}=
-\theta_{ba}= \theta (t_a,t_b)
\in [0,2\pi)$. $\A_{\G}$ is generated
by $U_a\equiv U_{t_a}$ satisfying
$U_a U_b= \exp
\{i\theta_{ab}\}U_b U_a$.
The dual $\A'_\G$ is generated by
$\Ut_a$'s, which satisfy
eqs. (\ref{18}) and (\ref{19}):
$\Ut_a U_b = U_b \Ut_a$ and
$\Ut_a \Ut_b = \exp
\{-i\theta_{ab}\}\Ut_b \Ut_a$.

These are just the quantum tori
introduced in ref. \cite{CDS}
in a different way. So
we are able to reproduce all
results there. In particular,
since $\Rt^\mu_\nu=\d^\mu_\nu$,
eq. (\ref{aU}) implies that
if we realize $\Ut_a$ as the
basic functions $\exp\{ i\sigma_a\}$
on the torus with coordinates
$0\le\sigma_a<2\pi$, then
$\dt^\mu$ are the derivatives
$-i d_a^\mu \del/\del \sigma_a$.
The noncommutative nature of the
torus is exhibited in the unusual
multiplication law for two
functions, pertinent to
eq. (\ref{19}):
\[(f_1\star f_2)(\sigma)
= \left.\exp\{\frac{i}{2}\theta_{ab}
{\del \over \del \sigma_a}
{\del \over \del \sigma'_b} \}
f_1(\sigma) f_2 (\sigma')
\right|_{\sigma=\sigma'}.\]
With $[X^\mu, X^\nu]$ understood
as $X^\mu\star X^\nu-X^\nu\star X^\mu$,
one gets a deformed Yang-Mills theory
parametrized by $\theta_{ab}$ on the
torus with coordinates $\sigma_a$
\cite{CDS}.

\subsection{Matrix Theory on ALE Orbifolds}

As the second example, let us consider
Matrix theory on ALE
orbifolds \cite{Doug,BCD}. One will see how the
results in type IIB theory \cite{DM,JM}
are recovered.

The ALE orbifolds are $\C^2/\G$,
where $\C^2$ is the complexification
of $\R^4$ by defining $Z^1=X^6+iX^7$
and $Z^2=X^8+iX^9$, and $\G$ is
a discrete subgroup of $SU(2)$
properly acting on $\C^2$.
Such subgroups have been classified
by Klein in last century \cite{Klein}.
They are all finite.
The action of $\G$ on $\C^2$ is homogeneous:
$\Phi_g(X)=R_g(X)$, where $R_g$ is the two
dimensional representation of $\G$ embedded
in the fundamental representation of $SU(2)$.

The solutions of the matrix variables are immediately
$Z^i=A^j(\Ut)\Rt^i_j$, $A_0=A_0(\Ut)$ and $X^\mu=X^\mu(\Ut)$
for $i, j=1,2$ and $\mu=1,\cdots,5$.

For the case of $\A_\G$ being the
untwisted group algebra $\C\G$,
$\H_\G$ is the genuine regular representation
of $\G$. The natural action of
$U_g$'s on ${\cal A}_\G$ are
represented by $|\G|\times |\G|$
matrices, which can be made
block-diagonal so that each irreducible
representation $R_i$ of $\G$ appears
as an $n_i\times n_i$ block $n_i$ times.
In the basis where $U_g$ are block-diagonal,
so are $A_0$ and $X^{\mu}$.
The gauge group $\Gc'$
is thus a product of unitary
groups for each block:
$F=\prod_{r} U(n_r N)$.
It can be shown that the operators
$\Rt$ are determined by the well-known
representation decomposition:
\be
\Phi\otimes R_r = \oplus_s a_{rs} R_s,
\label{decom}
\eq
where $a_{rs}$ are the elements of the
adjacency matrix $A$ of the simply
laced extended Dynkin diagrams. Namely,
$\Rt$'s connect the neighboring vertices
($R_r$ and $R_s$) in the extended
Dynkin diagram. So in the basis in which
the regular representation is
block-diagonal, $\Rt$'s consist of
off-diagonal blocks, which connect
neighboring unitary groups making up
the total gauge group $F$, with a
structure isomorphic to the adjacency
matrix $A$ for the extended Dynkin
diagram. (These considerations for
$\Rt$ can be generalized to projective
representations for arbitrary $\G$,
as $R_g$ is always a representation
of $\G$ even if $d^\mu_g\neq 0$.)

After taking into account of the fermionic
partners, we get hypermultiplets which
transform in the fundamental
representations of the unitary groups,
according to the representations
$\oplus a_{rs} ({\bf n}_r,{\bar{\bf n}}_s)$.
Pictorially, they correspond to the links
in the extended Dynkin diagram. Put
everything together, the field content
one obtains is the ${\cal N}=1, D=6$
supersymmetric Yang-Mills theory
dimensionally reduced to $0+1$ (or
$1+1$) dimensions, if we start with
the BFSS Matrix D0-brane (or string) theory.

\section{Orientifolds}

We may also consider actions of
$\G$ lifted to matrix variables other
than the natural one. For instance,
to extend our treatments to
incorporate orientifolds,
we need to consider a $\Z_2$-grading or
a $\Z_2$-graded extension of the group $\G$
\cite{GP}. This means that we associate a
number $n(g)=0,1$ to each element $g\in\G$
so that this assignment is compatible with
the product in the (extended) group:
\be
n(g)+n(h)\equiv n(gh)\quad (\mbox{mod} 2).
\eq
The quotient condition (\ref{qc}) for $U_g$
with $n(g)=0$ remains unchanged, while
for $n(g)=1$ it should be modified to \label{HWW}
\be
U_g^{-1} X^{\mu} U_g
= \Phi_g^{\mu}(X^T),
\eq
where $T$ denotes transposition of the matrices.
This is what we want for orientifolding, because
taking the transpose of the matrix variables
corresponds to reversing the orientitation of
the open strings connecting the D0-branes.

To put the quotient conditions for both $n(g)=0$
and $n(g)=1$ into the same form, instead
of $U_g$ we may consider $\Ub_g\equiv U_g C^{n(g)}$,
where $C$ is the complex conjugation operator.
Then it is not difficult to repeat the orbifold
construction above for the orientifolds
by including this $\Z_2$-grading.

In terms of $\Ub_g$, the quotient condition is
\be
\Ub_g^{-1}X^{\mu}\Ub_g=\Phi^{\mu}_g(X),
\eq
where the algebra of $\Ub_g$ is given by
\be
\Ub_g\Ub_h=e^{i\a(g,h)}\Ub_{gh}
\eq
for some $\a(g,h)$. If $n(g)=1$, then
$\Ub_g c=c^*\Ub_g$ for a complex number $c$.
The associativity of the algebra of $\Ub_g$
implies that
\bee
\d\a(f,g,h)&\equiv& (-1)^{n(f)}\a(g,h)-\a(fg,h)
+\a(f,gh)-\a(f,g)\\
&\equiv& 0 \quad (\mbox{mod} 2\pi);
\eqq
and shifting $\Ub_g$ by a phase factor $e^{i\b(g)}$
implies that
\be
\a(g,h)\rightarrow \a(g,h)-\d\b(g,h),
\eq
where $\d\b(g,h)\equiv (-1)^{n(g)}\b(h)-\b(gh)+\b(g)$.
The coboundary operator $\d$ defines a
cohomology $H^2(\G,U(1))$, which can be viewed as
the set of inequivalent consistent choices
of the algebra of $\Ub_g$ \cite{Ho}.
The equivalent classes of $\a$ in $H^2(\G,U(1))$ correspond
to possible backgrounds for the compactification.
Apparently the formulation of orbifolds can be viewed as
a special case of the orientifolds with the trivial
$\Z_2$-grading: $n(g)=0$ for all $g\in\G$.

Similarly we define the operators $\Ubt_g$ acting
on the Hilbert space spanned by $\Ub_g\ra$:
\be
\Ubt_g\Ub_h\ra=\Ub_h\Ub_g\ra,
\eq
and it follows that
\bee
&[\Ub_g,\Ubt_h]=0,\\
&\Ubt_g\Ubt_h=\Ubt_{hg}e^{i\a(h,g)\eps},
\eqq
where $\eps=(-1)^{\hat{n}}$ and
$\hat{n}\Ubt_g\ra=n(g)\Ubt_g\ra$.
For $(R_g)^{\mu}_{\nu}$, $d_g^{\mu}$ being real,
the solution of $X^{\mu}$ to the quotient condition is
\be
X^{\mu}=\left(A^{\nu}(\Ubt)(1-\hat{n})+
A^{\nu\ast}(\Ubt)\hat{n}\right)\Rt^{\mu}_{\nu}+\dt^{\mu},
\eq
where $\Rt^{\mu}_{\nu}$ and $\dt^{\mu}$ are
still defined by (\ref{Rtdt}), but now the projection
operator $P_g$ is defined by $P_g\Ub_h\ra=\d(g,h)\Ub_h\ra$.
Several examples of this general solution were presented
in ref. \cite{HWW}. Here the new insight provided by
the present treatment is that M(atrix) theory
compactified on orientifolds also corresponds to
noncommutative gauge theory.

\section{Presentation of the group $\G$}

In the above, we have worked with all
elements of $\G$; however, in
practice it may be more convenient to work
with a {\it presentation} (caution: not
representation!) of the discrete group $\G$.
By presentation we mean a finite set of
generators $g_a$'s $(a=1,2,...,r)$ and
a finite set of defining relations, $R:
f_m(g_1,...,g_r)=e,\; (1\le m\le k)$,
such that $\G$ is isomorphic to
the group $F$ freely generated by $g_a$
quotient by the equivalence relations $R$.
Then an arbitrary element $g$ of $\G$
can be written as a product of the
generators $g_a$, with the relations $R$
understood.

If a presentation of $\G$ is known,
we only need to write down the
quotient conditions for the generators,
with corresponding operators
$U_a\equiv U_{g_a}$. Also for a 2-cocycle,
we only need to introduce phase factors
for pairs of generators:
\be
U_aU_b = q_{a,b}U_{ab},
\qquad\quad |q_{a,b}| =1,
\label{2coc2}
\eq
or equivalently a phase factor for each
defining relation: each
$f_m(g_1,...,g_r)=e$ gives rise to
\be \label{df}
f_m(U_1,...,U_r)=p_m\,\one,
\qquad\quad |p_m| =1,
\eq
For instance, if $U_a$ commutes with
$U_b$, one may replace eq. (\ref{2coc2})
with
\be
U_a\, U_b =
\exp \{i\theta_{ab}\}\;U_b\, U_a,
\eq
where $\theta_{ab}$ is antisymmetric.
Working only with generators or with
defining relations simplifies the job
of finding all possible 2-cocycles.

We can generate the twisted group
algebra, ${\cal A}_{\Gamma}$, and the
representation Hilbert space,
${\cal H}_{\Gamma}$, in terms of the
generators $U_a$'s. Following
the above procedure, one can solve
the $X_i$'s in the quotient conditions
in terms of $\Ut_a$'s (or $\Ubt_a$'s),
which can be viewed as
a set of coordinates on the dual space.
Examples presented in ref. \cite{HWW}
were worked out explicitly in details in
this way.

So the use of presentation is
technically very helpful. However,
the presentation of a given group
$\Gamma$ may not be unique. We would
like to emphasize that the underlying
mathematics and physics are independent
of the choice of a presentation.
In particular it is possible that
different choices of generators in $\G$
can lead to essentially the same set
of ``deformed'' defining relations
(\ref{df}), when there is a corresponding
algebra automorphism on $\A'_\G$ (or $\A_\G$).
This can be understood as a global symmetry on the
dual space where the generators of $\A'_\G$
can be interpreted as coordinates.

\section{Discussions}

To conclude, the following remarks are
in order.

1) When the fermionic field $\Psi$ is taken
into account, the group $\G$ generically
will be extended into a larger group
acting on a superspace. Since the spinor
representation of spatial rotations is a
double covering of the vector representation,
the 2-cocycle $\a(g,h)$ may include the
operator $i\pi F$, where $F$ is the fermion
number operator. The Dirac operator acting
on $\Psi$ is then given by $\G^0 D_0+
\G^\mu X_\mu$ with $A_0$ and $X_\mu$ the
general solution to the quotient conditions.

2) In Matrix theory, there
should be many NCGT's resulting from
compactification on flat quotients
$\R^n/\G$, with
$\G$ being a point group
or space group in $\R^n$ (for $2\le n
\le 9)$ and allowing a nontrivial
2-cocycle.

3) Our approach can be easily used to
construct the quotient Matrix theory on
${\cal M}/\G$, if Matrix theory on
${\cal M}$ is known and has a discrete
symmetry $\G$ of ${\cal M}$. It also
applies to compactification of any
other matrix models, such as the IKKT
matrix model for IIB strings \cite{IKKT}.

\section{Acknowledgement}

P.M.H. thanks Yi-Yen Wu for discussion.
Work was supported in part by NSF
grant No. PHY-9601277 and a grant from
Monell Foundation.


\begin{thebibliography}{10}
\bibitem{comm1}
In string theory there should be no
pure gauge twist \cite{GP}.
\end{thebibliography}
\end{document}